\documentclass[12pt]{iopart}

\usepackage{graphicx}
\usepackage{amssymb}
\usepackage{wasysym}
\usepackage{lineno}

\linenumbers
\modulolinenumbers[10]

\begin{document}

\title{Spin incoherent transport in density-modulated quantum wires}

\author{$^1$K M Liu, $^1$H I Lin, $^2$V Umansky and $^1$S Y Hsu}

\address{$^1$ Department of Electrophysics, National Chiao Tung University, Hsinchu, 30010, Taiwan}
\address{$^2$ Braun Center for Submicron Research, Weizmann Institute of Science, Rehovot 76100, Israel}

\ead{syhsu@cc.nctu.edu.tw}

\begin{abstract}
Density, temperature and magnetic field dependences on electron transport in a quantum wire were studied. Decrease of carrier density gives a negative conductance correction on the first plateau at low temperatures. The prominent and mysterious ``0.7 structure" is more clearly resolved at low densities. The thermal behavior of the conductance follows the predictions of the spin-incoherent transport. The 0.7 structure at a low density drops to $e^2/h$ in a smaller in-plane magnetic field. The features are qualitatively consistent with the theoretical predictions and provide evidence for the existence of Wigner Crystal in low density quantum wires.
\end{abstract}


\section{Introduction}

In a 1D quantum wire the conductance is the integer multiple of $\frac{2e^2}{h}$ because of the quantized state in the transverse direction. Multiple plateaus of $i\cdotp(\frac{2e^2}{h})$ were mostly observed in experiments, where $i$ is an integer. When the conductance is close to $\frac{2e^2}{h}$, the electron density is effectively low. In this regime, coulomb interaction could become important, and the single particle model may not describe the physics satisfactorily. It has been long that an additional conductance shoulder was observed below the first plateau, and is often called ``0.7 anomaly/structure''. The very first attempting explanation referred it to spontaneous spin polarization.\cite{Thomas1996} Many works have been intrigued since then. It has been associated with spin polarization induced by exchange interaction\cite{Reilly2002}, thermal depopulation\cite{Kristensen2000} and Kondo-like effect due to unexpected magnetic impurities\cite{Cronnewett2002,Hirose2002,Hirose2003}. However, the mechanism is not yet fully understood for the time being.

  Recent theoretical and experimental works give further insight into this mystery. Electrons at low densities attempt to occupy equidistantly in order to minimize the coulomb repulsion and form a lattice-like structure, Wigner crystal.\cite{Wigner1934} It was proposed that an one dimensional Wigner crystal can be modeled as an `\textit{antiferromagnetic}' spin-chain with coupling strength J.\cite{Matveev2004a,Matveev2004b} The lattice vibration and spin interaction can be described by two separate Hamiltonians. Accordingly, charge-spin separation holds in a 1D Wigner crystal, which is a characteristic of Tomonaga-Luttinger Liquid(TLL).\cite{Haldane1981} The charge sector($\rho$) contributes a resistance of $\frac{h}{2e^2}$, $R_{\rho}$, while the resistance contributed from the spin sector($\sigma$) depends on the relative strength ratio between J and temperature T. When the spin sector is thermally excited, the excited part is reflected by the wire, giving $\displaystyle{R_{\sigma}=R_{0}e^{-\frac{\pi J}{2T}}}$. This was later termed as``spin-incoherent transport".\cite{Hew2008,Fiete2005} $R_0$ equals $\frac{h}{2e^2}$ in a `\textit{long}' TLL connected with non-interacting Fermi gas. The value of $R_0$ may depends on the wavelength of plasmons, i.e how strongly plasmons are aware of the interaction inside the wire. Therefore the conductance of an one-dimensional Wigner crystal depends on J, T and wire length. It has been reported that the conductance of a GaAs/AlGaAs quantum wire approaches $\frac{e^2}{h}$ with weakening confinement potentials and it was referred to spin-incoherent transport.\cite{Hew2008} Inspired by the aforementioned works, we studied the dependence of the conductance correction on carrier density, temperature and magnetic field in a semiconductor quantum wire. The results agree with the predictions of spin-incoherent theories and suggest connections with 0.7 structure.            

\section{Experimental Configurations}
The schematic layout of our sample is shown as the inset of Fig.\ref{fig1}(b). A pair of Ti/Au split gates(sg) is fabricated on the surface of GaAs/AlGaAs heterostructure with the 2-Dimensional-Electron-Gas(2DEG) located 90nm beneath.The mobility and carrier density in the dark of the 2DEG are $\sim 2.2 \times 10^6 cm^2 / V \cdot s $ and $\sim 1.4 \times 10^{11} cm^{-2}$, respectively. On each cooldown, infrared was employed to excite the device and the carrier density rises to $\sim 2.4 \times 10^{11} cm^{-2}$ determined by Shubnikov-de Haas oscillations. The mean free path of 2DEG is more than $10\mu m$. The physical dimensions of the slit of the split gates are $0.4{\mu}m\times 0.4{\mu}m$ in width and length. A quasi-1D quantum wire is created by the depletion of 2DEG beneath the negatively biased split gates. On top of them, being isolated by a dielectric layer of cross-linked Polymethylmethacrylate(PMMA), a top gate(tp) is used to control the electron density, $n_{w}$. The device fabrication is accomplished by general electron-beam lithography and lift-off technique. Measurement was carried out mainly in a $^3$He cryostat and also in a dilution refrigerator. Base temperatures are 270 and 30mK, respectively. Four-terminal differential conductance was recorded by standard lock-in techniques. The amplitude of source-drain ac excitation ranges 2-10$\mu V$, with the frequency in a range of 13-51Hz. The results presented in this paper were done on numerous cooldowns and are reproducible.

\section{Results and Discussion}
\begin{figure}[h]
\centering
\begin{minipage}[h]{75mm}
\centering
\includegraphics[scale=0.75]{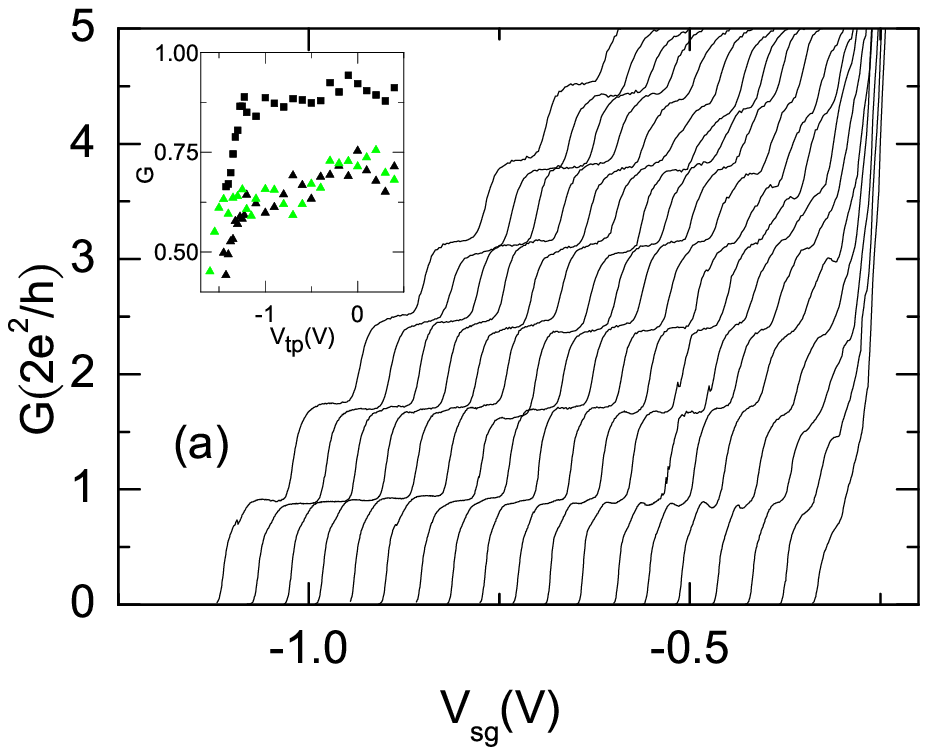}
\end{minipage}
\begin{minipage}[h]{75mm}
\centering
\includegraphics[scale=0.75]{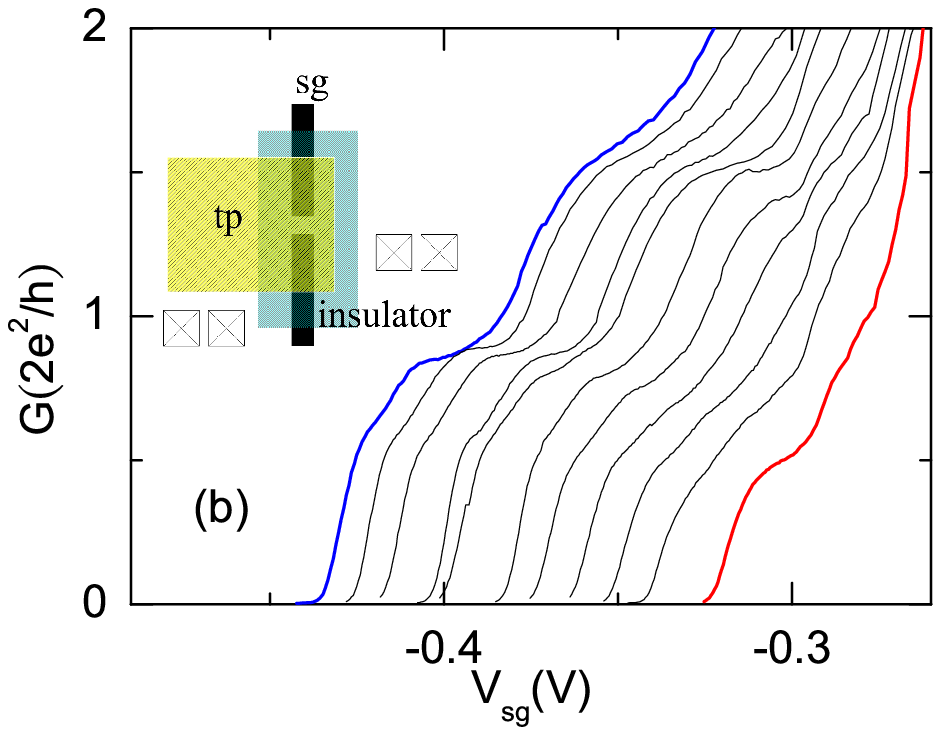}
\end{minipage}
\caption{(a) Conductance as a function of the split gate voltage $V_{sg}$, against the top gate voltage, $V_{tp}$. From left to right: $V_{tp}=+0.4$ to $-1.4V$ in steps of 0.1V at T=0.27K. Inset: Conductance features vs. $V_{tp}$ on the first plateau($\blacksquare$) and the 0.7 structure($\blacktriangle$) for two samples at 0.27K. (b) G vs. $V_{sg}$ in finer steps of $V_{tp}$. Thick blue: $V_{tp}=-1.2V$; thick red: $V_{tp}=-1.45V$ Inset: The schematic layout of the sample. }
\label{fig1}
\end{figure}

Firstly, we discuss the effect of carrier density. Fig.\ref{fig1}(a) shows conductance as a function of split gate voltage $V_{sg}$ against top gate voltages, $V_{tp}$ at 0.27K. From left to right, $V_{tp}$ decreases from +0.4 to -1.4V. Conductance quantizations are observed. The number of `integer' plateaus reduces from eleven to one, indicating that electron density reduces with respect to decreasing $V_{tp}$. The 0.7 structure is weakly perceptible for $V_{tp}\gtrsim-0.4V$. On the other hand, 0.7 structure is better resolved at lower densities for $V_{tp}\lesssim-0.4V$. Notice that two clear steps are recognizable in the traces of $-0.6$ to $-1.0V$. Fig.\ref{fig1}(b) shows a closer view of the low density regime. Two plateaus standing for the first subband and 0.7 structre have conductances of 0.85 and 0.64 respectively at $V_{tp}=-1.2V$. Conductances of both structures decrease with further reducing $V_{tp}$. The first plateau finally {\it recedes} at $V_{tp}=-1.45V$. A plateau of 0.49$(\frac{2e^2}{h})$ {\it lasts} which {\it evolves from 0.7 structure} at higher densities.

The positions of the first plateau and 0.7 structure are located by differentiating the traces and finding the minimums nearby the pinch-off regime. The overall correlation between the conductances and $V_{tp}$ on the first plateau and 0.7 structure is organized and shown in the inset of Fig.\ref{fig1}(a) The conductance of the first plateau decreases from 0.91 at high densities to 0.66 at low densities while 0.7 structure decreases from 0.71 to 0.49. Resembling features are also observed in the second sample demonstrated as green triangles. In the model of 1D Wigner crystal, the coupling strength J decreases with reducing $n_{w}$, because the coulomb barrier is boosting. Thus, the portion of incoherent spin swells as well as the \textit{serial resistance} $R_{\sigma}$, agreeing with the downward trend of \textit{conductance} with respect to $V_{tp}$. Both traces of first plateau and 0.7 structure have small slopes from $V_{tp}=$+0.4 to -1V, i.e the conductance decrease slowly, but drops rapidly for $V_{tp}\lesssim -1V$. There seems to be a critical density that dramatically manifest features of spin-incoherent transport. 

One would be concerned that whether the conductance decrease is dominated by the serial resistance contributed by the 2DEG below the top gate area, $R_{2DEG}$.  It is justified not being the case. By fixing the lower split gate at -0.7V and sweeping $V_{tp}$(not shown), $R_{2DEG}$ increases from $\sim 90\Omega$ to $\sim 400\Omega$ for $V_{tp}=-1V$. Nevertheless, a conductance reduction from 0.71 to 0.6$(\frac{2e^2}{h})$ in the same range represents a resistance difference $\sim$3.5K$\Omega$, much larger than the increment of $R_{2DEG}$.

\begin{figure}[h]
\centering
\begin{minipage}[h]{50mm}
\centering
\includegraphics[scale=0.75]{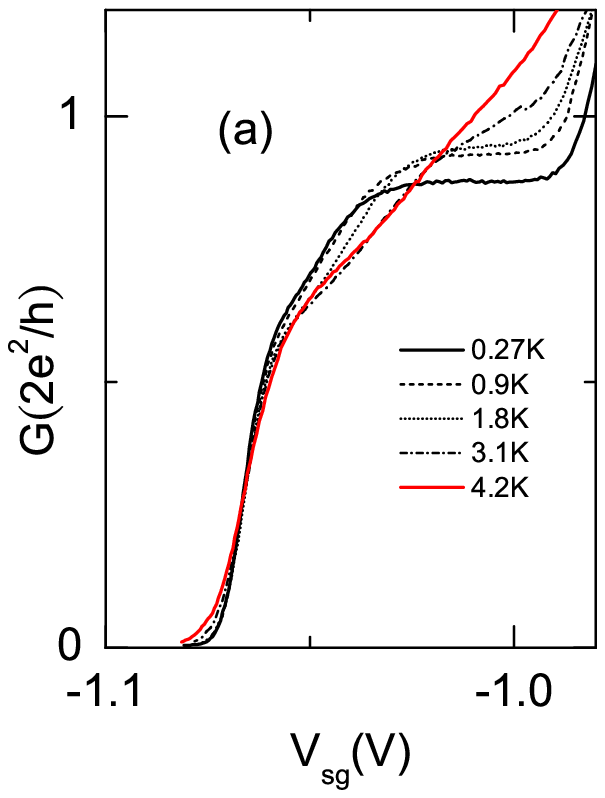}
\end{minipage}
\begin{minipage}[h]{50mm}
\centering
\includegraphics[scale=0.75]{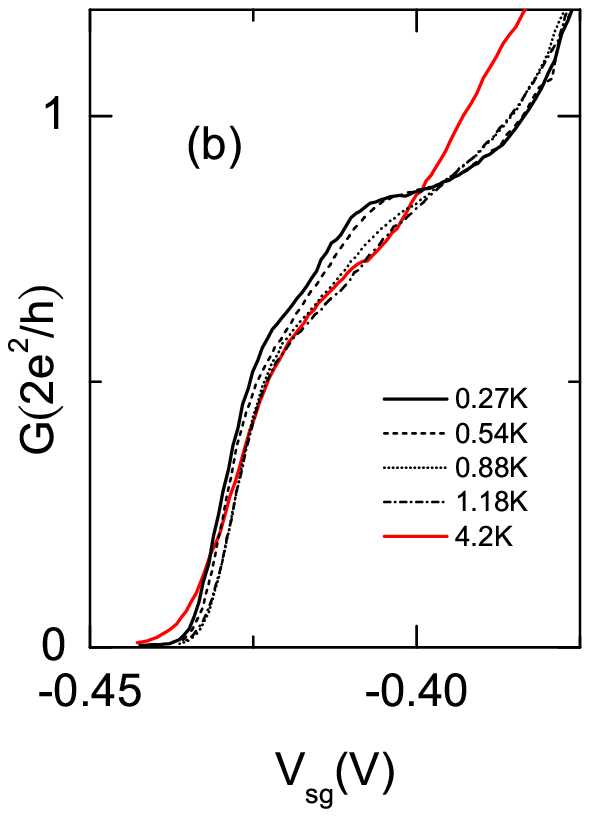}
\end{minipage}\begin{minipage}[h]{50mm}
\centering
\includegraphics[scale=0.75]{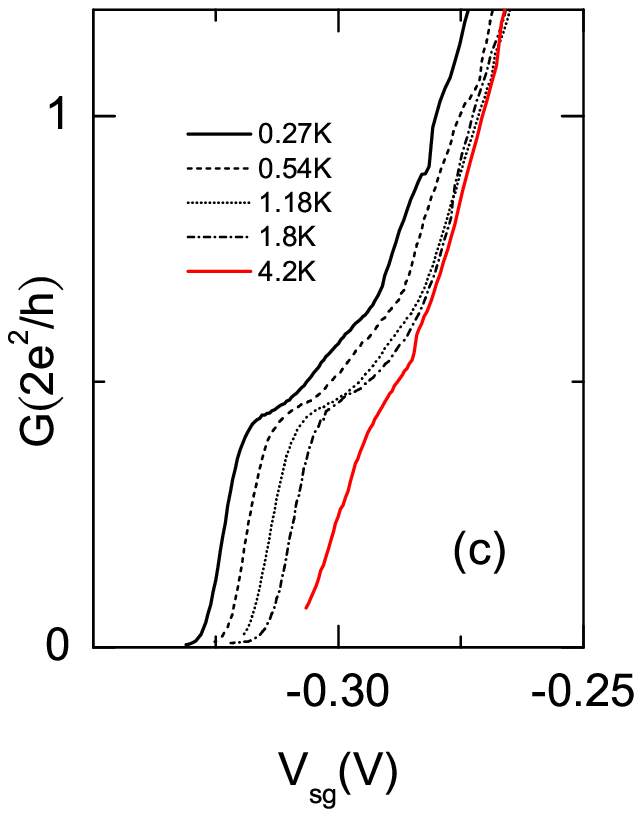}
\end{minipage}
\caption{G($V_{sg}$) as a function of temperature at various $V_{tp}$.(a) $V_{tp}$=+0.3V. (b)$V_{tp}$=-1.2V. (c) $V_{tp}$=-1.45V. The traces are shifted for clarity.}
\label{fig2}
\end{figure}

Next, we study the effect of temperature. Similar studies were reported in previous works with somewhat different device designs\cite{Kristensen2000,Thomas2000}, however we found rather diversified temperature dependence at different densities. Fig.\ref{fig2} shows G($V_{sg}$) at various temperatures for $V_{tp}=$ +0.3, -1.2, and -1.45V. In Fig.\ref{fig2}(a) and \ref{fig2}(b), 0.7 structure is robust against temperatures up to 4.2K. Surprisingly, the critical temperatures that the first plateau is suppressed by thermal smearing, remaining the 0.7 structure are $\sim$3.1 and 1.18K, respectively. In Fig.\ref{fig2}(c), the 0.49$(\frac{2e^2}{h})$ plateau remains nearly unchanged up to 1.8 from 0.27K. The manifoldly thermal behavior implies various J for different $V_{tp}$. $J$ appears to decrease with decreasing $V_{tp}$, hence leading to the reduction of critical temperature down to $<$0.27K at $V_{tp}=-1.45$V. A small J at $V_{tp}=-1.45$V also suggests that phonon scatterings can take over at lower temperatures. As a result, the plateau is washed out by further increasing the temperature to 4.2K, in contrast to the other two cases.

\begin{figure}[h]
\centering
\begin{minipage}[h]{50mm}
\centering
\includegraphics[scale=0.75, trim=5mm 0mm -5mm 0mm]{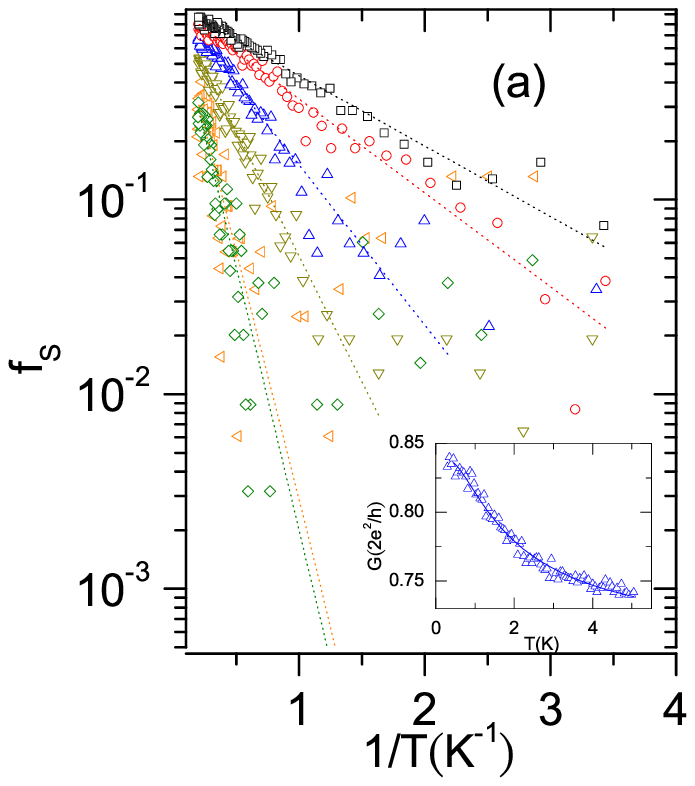}
\end{minipage}
\centering
\begin{minipage}[h]{50mm}
\centering
\includegraphics[scale=0.75,trim=3mm 0mm -7mm 0mm]{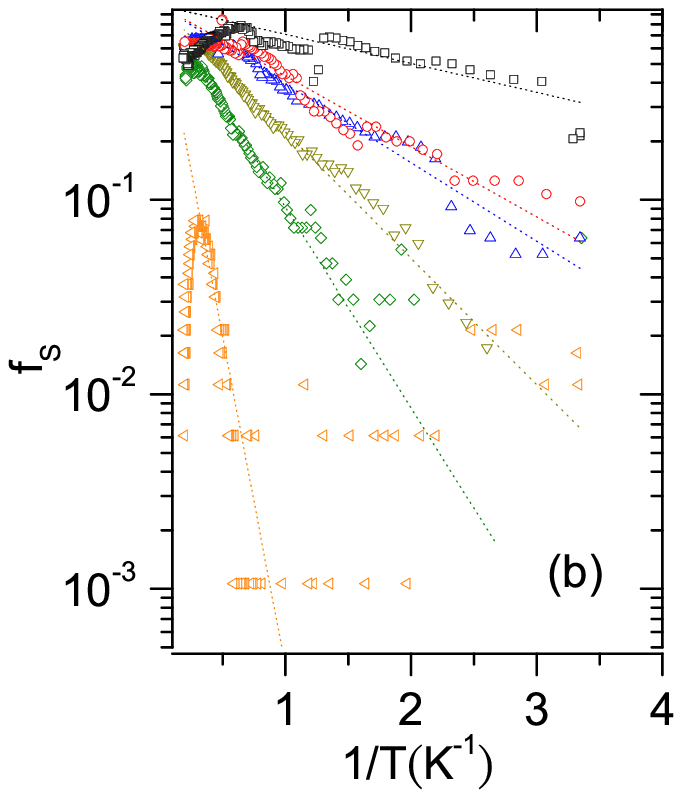}
\end{minipage}
\begin{minipage}[c]{80mm}
\centering
\includegraphics[scale=0.75,trim=5mm 0mm 5mm 10mm]{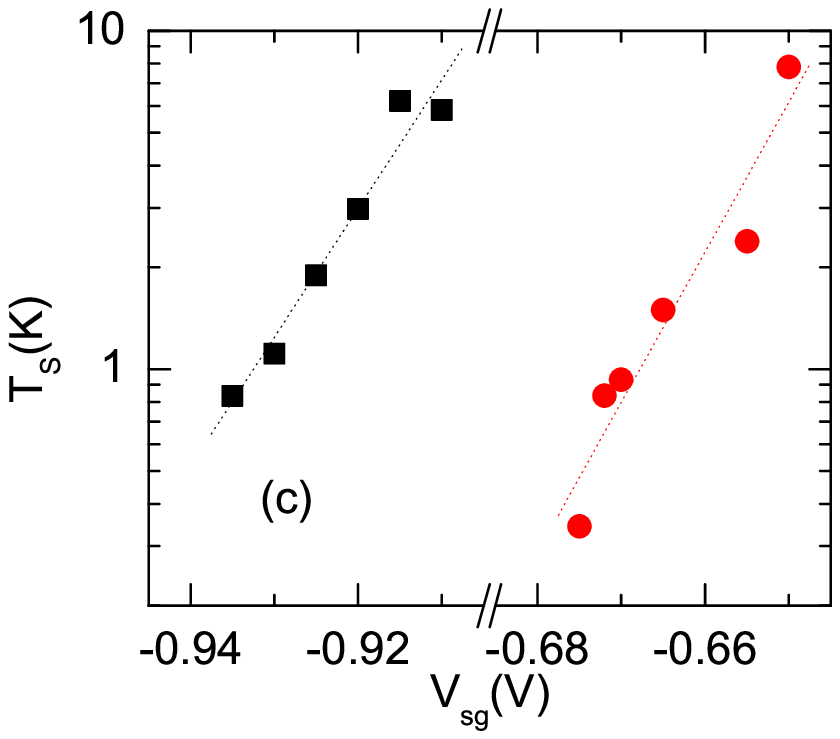}
\end{minipage}
\caption{Semilogarithmic plot of $f_{S}$(see text) vs. inverse temperature  $\frac{1}{T}$ at various locations between 0.7 structure and the first plateau. Points are experimental results whence lines are linear fittings. (a) $V_{tp}=0V$. Traces are $g_{0}=$ 0.912($\vartriangleleft$), 0.887($\Diamond$), 0.867($\triangledown$), 0.839($\vartriangle$), 0.794($\ocircle$) and 0.743($\square$) at 0.27K. Inset: The experimental result and numerical fitting of G(T) for $g_{0}=$0.794. (b) $V_{tp}=-0.5V$. $g_{0}=$0.893($\vartriangleleft$), 0.884($\Diamond$), 0.858($\triangledown$), 0.810($\vartriangle$), 0.764($\ocircle$) and 0.732($\square$).(c) $T_{S}$ extracted from the numerical fittings as a function of the corresponding $V_{sg}$ for $V_{tp}=0$($\blacksquare$) and $-0.5V$($\CIRCLE$).}  
\label{fig3}
\end{figure}

The trace of points in the inset of Fig.3(a) is a typical result of G(T) at a fixed electron density (with fixed $V_{tp}$ and $V_{sg}$) . As expected for the evolution of the 0.7 structure, conductance declines with increasing temperature. In order to compare with the theoretical predictions, we fit the data with a modified formula,
\begin{equation}
G_{w}(T)=\displaystyle{g_{0}\frac{2e^2}{h}(1+\delta r \cdotp e^{- T_{S}/T})^{-1}}
\end{equation}
Rearranging Eq.(1), f$_{S}$ defined as $\displaystyle{\frac{1}{\delta r}(\frac{g_{0}\frac{2e^2}{h}}{G(T)}-1)}$ is equal to $e^{-\frac{T_{S}}{T}}$ . Semilogarithmic plots of f$_S$ versus 1/T are presented to reveal such typical activation behavior. Figs.\ref{fig3}(a) and \ref{fig3}(b) summarize data and the fit results at various electron densities. It is important to point out that the Kondo-like model could not be fitted to our data. There are two sets of traces for different $V_{tp}$ in Fig.\ref{fig3}. In Fig.\ref{fig3}(a), $V_{tp}$=0V corresponding to a high electron density and $g_{0}$ spans from 0.91 to 0.74 at 0.27K. As seen, all traces generally follow the linear behavior in this semilogarithmic Arrhenius plot. Such linear relation holds in the temperature range shown here, 5-0.27K, for the topmost trace with $g_{0}$=0.74. The slope of the linear fit represents $T_S$. $T_S$ increases while the temperature range of linear relation shrinks with increasing $g_{0}$. G(T) deviates from the activation form and seems to saturate at low temperatures. For an extreme case, the temperature range is from 5K to 1.5K and the conductance increases slowly around 1.5K and decreases at lower temperatures for green trace with $g_{0}$=0.89. Such deviations have been reported in other similar systems\cite{Sfigakis2008,Hew2008}. The proportion of deviation is greater with a larger $g_{0}$. This feature is possibly due to weak localization in a quantum wire, since impurities may manifest themselves as localization effects in a ballistically traveling electronic system(liquid) instead of a crystallized one(Wigner crystal). 
At low carrier densities(smaller $g_{0}$), the electronic phase is more likely to be the Wigner crystal. 

The data of $V_{tp}=-0.5V$ in Fig.\ref{fig3}(b) exhibits similar characteristics. The traces have smaller slopes, indicating smaller activation temperatures. Comparing Eq.(1) with the theoretical form of $G_{w}$, $T_S$ is indeed the coupling strength J multiplied with a constant, $\pi$/2. Fig.\ref{fig3}(c) is a semilogarithmic plot of $T_S$ as a function of the split gate voltage for sets of traces of $V_{tp}=$0 and -0.5V. Since both $V_{tp}$ and effectively $V_{sg}$ control the electron density $n_w$, $T_S$ decreases with decreasing either $V_{tp}$ or $V_{sg}$ implying that J decreases with reducing $n_w$. This scenario supports 1D Wigner crystal model that J is determined by tunneling through the density dependent Coulomb barrier. That J$\sim n^2e^{-1/\sqrt{n}}$ is obtained by Matveev\cite{Matveev2004b}. Although the value of $n_w$ is not obtained for our data sets of different $V_{tp}$ and $V_{sg}$, that J depends exponentially on $V_{sg}$ shown in Fig.\ref{fig3}(c) and n($V_{sg}$) from other findings\cite{Hamilton1992} make the experimental results a step closer to the theoretical prediction. In our results, J ranges from 0.3 to 5K in consistence with the condition, T$\sim$J, for thermally excited spin sector at low electron densities.

$\delta r$ characterizes the resistance constant $R_{0}$ due to the spin-incoherent part in the quantum wire. The average $\bar{\delta r}$ over six traces for each top gate voltage in Fig.\ref{fig3}(a) and \ref{fig3}(b) is 0.169 and 0.186 respectively. $\bar{\delta r}$ is less than one fifth instead of {\it one} which is the theoretical prediction. The reason for the discrepancy may be that the wire length in our experiment is short compared with the theoretical assumption.\cite{Matveev2004a} Coupling between the interacting quantum wire and the noninteracting Fermi contacts suppresses the TLL properties.\cite{Maslov1995,Ponomarenko1995,Safi1995} Moreover, $\bar{\delta r}$ has no significant difference between top gate voltages. This is sensible since the device geometry is invariable. The temperature dependence reported in Ref.\cite{Kristensen2000} matches with the equation, ${G(T)=g_{0}\frac{2e^2}{h}(1-C\cdotp e^{-T_{S}/T})}$. Notice that this is an approximation of Eq.(1), since for small $\delta r \cdotp e^{-T_{S}/T}$, it can be expanded up to the linear term and have the same formula.

\begin{figure}[h]
\centering
\begin{minipage}[h]{50mm}
\centering
\includegraphics[scale=0.75]{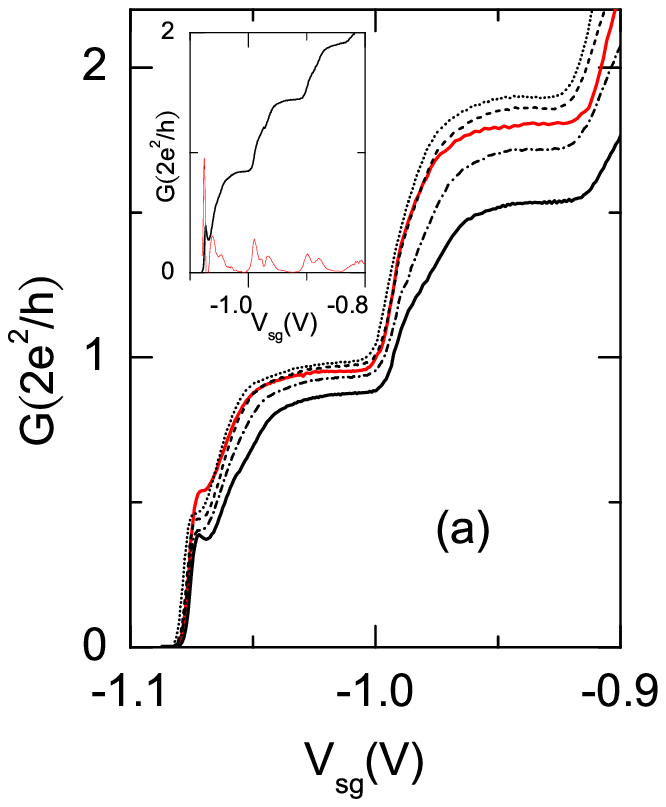}
\end{minipage}
\centering
\begin{minipage}[h]{50mm}
\centering
\includegraphics[scale=0.75]{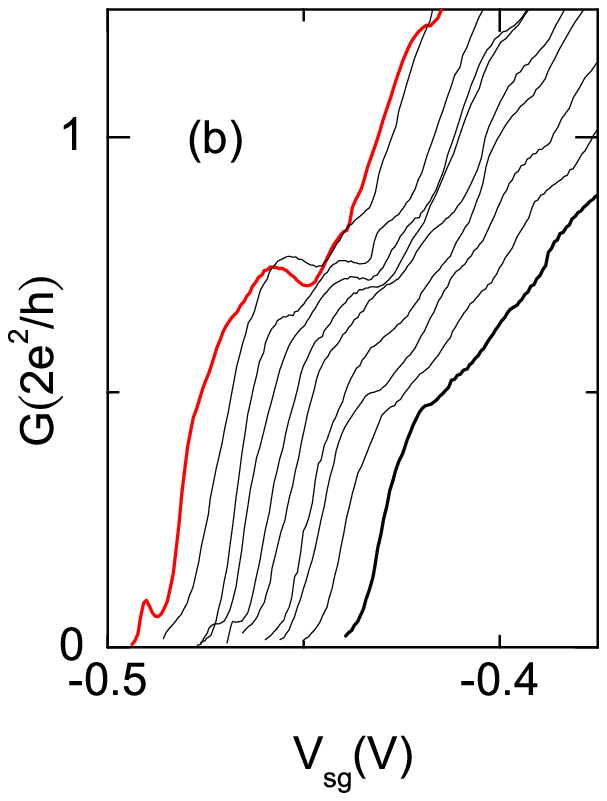}
\end{minipage}

\caption{G vs. $V_{sg}$ as a function of the in-plane magnetic fields perpendicular to the quantum wire, $B_{\perp}$, at 100mK. (a)$V_{tp}=0V$. $B_{\perp}=$0(thick red), 2({\tiny\dotted}), 4({\tiny\dashed}), 6({\tiny\chain}) and 8T(thick black). Inset: G vs. $V_{sg}$ at $B_{\perp}=9T$(black) along with its derivative in arbitrary unit(red). (b)$V_{tp}=-1.4V$. $B_{\perp}=$0(thick red) to 9T(thick black) in steps of 1T.  }
\label{fig4}
\end{figure}

Finally, we study the effect of an in-plane magnetic field $B_{\perp}$ applied perpendicularly to the quantum wire. Since the 1D Wigner crystal is viewed as an antiferromanetic spin-chain, the field that fully polarizes the quantum wire depends on coupling strength J. When the quantum wire is fully polarized, in other words--spin degeneracy is lifted, one spin mode contributes $\frac{e^2}{h}$ to the conductance. We first present G($V_{sg}$) for $V_{tp}=0$ in Fig.\ref{fig4}(a). The 0.7 structure is not perceived at $B_{\perp}=0$. Increasing the magnetic field to 8T, the conductance of the first plateau decreases from 0.953 to 0.870. On the other hand, the plateau of $\frac{e^2}{h}$ just begins to develop at $B_{\perp}=8T$. Plateau splitting is better recognized in the inset of Fig.\ref{fig4}(a). The half-plateau associating with spin splitting does not fully develop in 9T, however, $dG/dV_{sg}$ shows clearly double-peak structure. It was reported that the 0.7 structure evolves to the spin-polarized plateau.\cite{Thomas1996} The results of $V_{tp}=-1.4V$, which is expected to have a smaller  J, is presented in Fig.\ref{fig4}(b). The conductance of the `\textit{anomaly}' is about 0.746 in 0T. By increasing the magnetic field, the conductance falls continuously, reaches $\sim$0.491 in $B_{\perp}=7T$ and {\it saturates} thereafter. The Zeeman energy $\epsilon_{B}=|g^{*}| \mu_{B}B$ is $\sim 178.6\mu eV$ at 7T. The estimated $J=\frac{\epsilon_{B}}{2}$ is about 1.04K being comparable with the extracted $T_{S}$ in the previous discussions.

Admittedly, one would be concerned about the relation between $J$ and carrier density $n_{w}$. After all, the precise density of a quantum wire is not easy to know.\cite{Berggren1986} We regardlessly strove to roughly estimate the 1D density, $n_w=\sqrt{\frac{8mE_{F}^{*}}{\pi^2 \hbar^2}}$, with the first subband populated. The density and energy $E_0$ at the pinch-off is assumed to be zero, while the effective Fermi energy is rendered as $E_{F}^{*}=E_0+\Delta E_{1,2}$. The level spacing  between the first two subbands, $\Delta E_{1,2}$, extracted from the transconductance spectroscopy(not shown) is $\sim$1.5meV for $V_{tp}=0$. It gives a density of $\sim 3.27 \times 10^{7} m^{-1}$. According to Ref.\cite{Matveev2004a}, this density suggests $J\sim$0.52K using the Bohr radius of GaAs $a_{B}\sim$10.34nm, being smaller but not far away from those values obtained by temperature and magnetic field dependences. However, the precise relation between $J$ and $n_{w}$ requires further investigation.

\section{Summary}
In summary, we have studied the influences of carrier density, temperature and magnetic field on the conductance close to the first plateau in quantum wires. A negative conductance correction increases with the decreasing carrier density. This effects leads to the well resolved 0.7 structure at low densities which is weakly observable in the opposite regime. The conductance of the 0.7 structure drops to $\frac{e^2}{h}$ at low densities. The conductance as a function of the temperature follows the formula deduced from the spin-incoherent model, showing thermal activated behaviors. The activated temperature {\it exponentially} depends on the split gate voltage. As for the magnetic dependence, the conductance reduces to $\frac{e^2}{h}$ more rapidly at a lower density with respect to an increasing in-plane magnetic field. The results are consistent with the predictions based upon the picture of 1D Wigner crystal. We remark that non-interacting single particle theories are not sufficient to explain the results of the explored regime, low electron densities particularly. Many-body effects manifest in a semiconductor quantum wire and Tomonaga-Luttinger Liquid is likely to be the ground state.


\ack
This work was supported by NSC grant in Taiwan under project No NSC96-2112-M-009-030-MY3 and MOE ATU program. We thank for precious discussions with K. A. Matveev.

\end{document}